\documentclass[mathleft
]{an}
\usepackage{graphicx}
\usepackage{times}
\begin{document}

% The following seven commands are intended for editorial usage and should be ignored by
% the author(s).
\Pagespan{789}{}% Document's page range.
% If second parameter is left empty, the last page is computed automatically.
\Yearpublication{2013}%
\Yearsubmission{2013}%
\Month{12}%
\Volume{999}%
\Issue{88}%
% \DOI{This.is/not.aDOI}%

\title{Rotational Behaviors and Magnetic Field Evolution of Radio Pulsars}

\author{Yi Xie \inst{1,2}
%Example
%for footnote, note the usage of the \texttt{fnmsep}
%command as separator between institute number and footnote mark}
\and  Shuang-Nan Zhang \inst{1,3,4} \fnmsep \thanks{Corresponding
author: \email{zhangsn@ihep.ac.cn}}}

\titlerunning{Rotational Behaviors and Magnetic Field Evolution of Radio Pulsars}
\authorrunning{Yi Xie \& Shuang-Nan Zhang}

\institute{National Astronomical Observatories, Chinese Academy of
Sciences, Beijing, 100012, China \and University of Chinese Academy
of Sciences, Beijing, 100049, China \and Key Laboratory of Particle
Astrophysics, Institute of High Energy Physics, Chinese Academy of
Sciences, Beijing 100049, China \and Physics Department, University of Alabama in
Huntsville, Huntsville, AL 35899, USA}

\received{30 Dec 2013} \accepted{30 Dec 2013} \publonline{later}


\abstract{ The observed long-term spin-down evolution of isolated
radio pulsars cannot be explained by the standard magnetic dipole
radiation with a constant braking torque. However how and why the
torque varies still remains controversial, which is an outstanding
problem in our understanding of neutron stars. We have constructed a
phenomenological model of the evolution of surface magnetic fields
of pulsars, which contains a long-term decay modulated by short-term
oscillations; a pulsar's spin is thus modified by its magnetic field evolution.
The predictions of this model agree with the precisely measured spin evolutions of several individual pulsars;
the derived parameters suggest that the Hall drift and Hall waves in the NS
crusts are probably responsible for the long-term change and
short-term quasi-periodical oscillations, respectively. Many statistical properties of the timing noise of pulsars can be well re-produced with this model, including correlations and the distributions of the observed braking indices of the pulsars, which
span over a range of more than 100 millions. We have also
presented a phenomenological model for the recovery processes of
classical and slow glitches, which can successfully model the
observed slow and classical glitch events without biases.}

\maketitle

\section{Introduction}
A radio pulsar is a rotating neutron star (NS) with strong surface
magnetic fields, which emits a beam of electromagnetic radiation
along the axis of the fields. Much resembling the way of a
lighthouse, the radiation can only be seen when the light is pointed
to the direction of an observer. Since a NS is a very stable
rotator, it produces a series of pulses with a very precise interval
that ranges from milliseconds to seconds in the radio band.

The arrival times of the pulses can be recorded with very high
precision. Indeed, thanks to the high precision, a surprising amount
can be learned from them (see Lyne \& Graham-Smith 2012). As early
as the first pulsar was discovered (on November 28, 1967), Hewish
and his collaborators noticed that they provide the information that
not only the radio source might be a rotating NS, but also about its
position and motion, as well as the dispersion effect during the
pulses' propagation through the interstellar medium (Hewish el al.
1968).

The time-of-arrival (TOA) measurements now give the precise
information on various modes of the spin evolution of individual
NSs, and on the orbits and rotational slowdown of binary pulsars,
and thus made possible some fundamental tests of general relativity
and gravitational radiation. Some millisecond pulsars with rather
stable pulsations (even challenging the best atomic clocks), are
used as a system of Galactic clocks for ephemeris time or
gravitational wave detection.

The variations of spin frequency $\nu$ and its first derivative
$\dot{\nu}$ of pulsars are obtained from polynomial fit results of
arriving time epochs (i.e. phase sequences) of pulses. Since the
rotational period is nearly constant, these observable quantities,
$\nu$, $\dot{\nu}$ and $\ddot{\nu}$ can be obtained by fitting the
phases to the third order of its Taylor expansion over a time span
$t_{\rm s}$,
\begin{equation}\label{phase}
\Phi_i = {\Phi} + \nu (t_i-t) + \frac{1}{2}\dot \nu (t_i-t)^2 +
\frac{1}{6}\ddot\nu (t_i-t)^3.
\end{equation}
One can thus get the values of $\nu$, $\dot{\nu}$ and $\ddot{\nu}$
at $t$ from fitting to Equation~(1) for independent $N$ data blocks
around $t$, i.e. $i=1,...,N$.

The most obvious feature of a pulsar spin evolution is that it is
observed to slow down gradually, i.e. $\dot\nu<0$. According to
classical electrodynamics, a inclined magnetic dipole in vacuum lose
its rotational energy via emitting low-frequency electromagnetic
radiation. Assuming the pure magnetic dipole radiation as the
braking mechanism (e.g. Lorimer 2004), we have
\begin{equation}\label{braking law}
\dot\nu =-A B_0^2 \nu^3,
\end{equation}
in which $A=8\pi^2R^6\sin\theta^2/3c^3I$ is a constant, $B_0$ is the
strength of the dipole magnetic fields at the surface of the NS,
$R~(\simeq10^6~{\rm cm})$, $I~(\simeq10^{45}~{\rm g~cm^2})$, and
$\theta~(\simeq\pi/2)$ are the radius, moment of inertia, and angle
of magnetic inclination from the rotation axis, respectively. Though
the energy flow from a pulsar may be a combination of this dipole
radiation and an outflow of particles, Equation (\ref{braking law})
is still valid, since the magnetic energy dominate the total energy
of the outer magnetosphere (Lyne \& Graham-Smith 2012).
\begin{figure}[h]
\center{
\includegraphics[angle=0,scale=0.3]{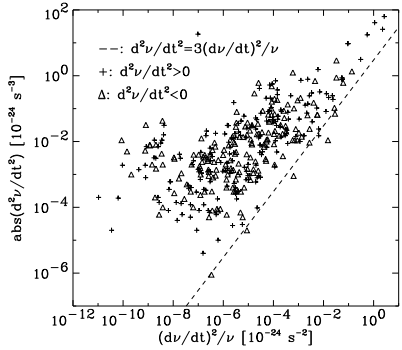}}
\caption{Observed correlation between $\ddot \nu$ and
$3\dot{\nu}^2/\nu$. The prediction of the standard magnetic dipole
radiation model, i.e., $\ddot \nu=3\dot{\nu}^2/\nu$ is shown as the
dashed line, which under-predicts significantly the magnitudes of
$\ddot \nu$ for most pulsars and also cannot explain $\ddot \nu<0$
for nearly half of the pulsars. All the data are taken from Hobbs et
al. (2010), and the figure is taken from Paper I.} \label{Fig:1}
\end{figure}

Following Equation (\ref{braking law}) and assuming $\dot B_0=0$,
the frequency's second derivative can be simply expressed as
\begin{equation}\label{ddotnu}
\ddot\nu=3\dot\nu^2/\nu.
\end{equation}
The model predicts $\ddot\nu>0$ and $|\ddot\nu|$ should be very
small. However, as shown in Figure \ref{Fig:1} (Zhang
\& Xie 2012a; hereafter Paper I), the observed
$\ddot{\nu}$ is often significantly different from the model
predictions, so that the braking mechanism may be oversimplified.
However how and why the torque varies still remains controversial,
which is an outstanding problem in our understanding of neutron
stars.

In this paper, we give a brief review for the phenomenological model
we constructed recently, and its applications on the spin behaviors
of pulsars, which include the statistical properties of $\ddot\nu$
and $n_{\rm b}$, and the dipole magnetic field evolution of some
individual pulsars, as well as their physical implications on NS
interiors. A phenomenological model for glitch recoveries of
individual pulsars are also briefly discussed.

\section{The phenomenological model}

To model the discrepancy between the observed $\ddot\nu$ and the
predicted $\ddot\nu$ by Equation (\ref{ddotnu}), the braking law of a pulsar is generally
assumed as
\begin{equation}
\dot \nu  =-K\nu ^{n_{\rm b}},\label{braking_law}
\end{equation}
where $n_{\rm b}$ is called its braking index. Manchester \& Taylor
(1977) gave that
\begin{equation}
n_{\rm b}=\ddot{\nu}\nu/\dot{\nu}^2,\label{braking_index}
\end{equation}
if $\dot{K}=0$. For the standard magnetic dipole radiation model
with constant magnetic field ($\dot{K}=0$), Equation (\ref{ddotnu}) applies and yields $n_{\rm b}=3$. Therefore
$n_{\rm b}\ne 3$ indicates some deviation from the standard magnetic
dipole radiation model with constant magnetic fields.

Blandford \& Romani (1988) re-formulated the braking law of a pulsar
as,
\begin{equation}
\dot \nu  =-K(t)\nu ^3.\label{blandford1}
\end{equation}
This means that the standard magnetic dipole radiation is
responsible for the instantaneous spin-down of a pulsar, but the
braking torque determined by $K(t)$ may be variable. In this
formulation, $n_{\rm b}\ne 3$ does not indicate deviation from the
standard magnetic dipole radiation model, but means only that $K(t)$
is time dependent. Assuming that magnetic field evolution is
responsible for the variation of $K(t)$, we have $K=AB(t)^2$, in
which $B(t)$ is the time variable dipole magnetic field strength of
a pulsar. The above equation then suggests that $n_{\rm b}< 3$
indicates magnetic field growth of a pulsar, and vice versa, since
$\dot{\nu}<0$ and $K>0$. This can be seen more clearly from (Paper I and Zhang
\& Xie 2012b (Paper II)),
\begin{equation}
\dot{K}=\frac{\dot{\nu}^2}{\nu^4}(3-n_{\rm b}).\label{zhang2012}
\end{equation}

Equation (\ref{blandford1}) cab be re-written as
\begin{equation}\label{braking law2}
\dot\nu \nu^{-3}=-A B(t)^2,
\end{equation}
and the time
scale of the magnetic field long-term evolution of each pulsar (see
Equation~(6) in Paper I) is given by
\begin{equation}\label{tau_B}
\tau_{B}\equiv \frac{B}{\dot B}=\frac{2\dot\nu_0\nu_0}{\ddot\nu_{\rm
L}\nu_0 -3\dot\nu_0^2}.
\end{equation}
$\tau_{B}<0$ indicates magnetic field decrease and
vice versa.

In Paper I and II, we constructed a phenomenological model for the
dipole magnetic field evolution of pulsars with a long-term decay
modulated by short-term oscillations,
\begin{equation}\label{B evolution}
B(t)=B_{\rm d}(t)(1+\sum k_i\sin(\phi_i+2\pi\frac{t}{T_i})),
\end{equation}
where $t$ is the pulsar's age, and $k_i$, $\phi_i$, $T_i$ are the
amplitude, phase and period of the $i$-th oscillating component,
respectively. $B_{\rm d}(t)=B_0(t/t_0)^{-\alpha}$, in which $B_0$ is the
field strength at the age $t_0$, and $\alpha$ is the power law
index.

By substituting Equation (\ref{B evolution}) into Equation
(\ref{braking law2}) and taking only the dominating oscillating
component, we obtained the analytic approximation for $\dot\nu$ (Xie
\& Zhang 2013c, hereafter Paper V):
\begin{equation}\label{vdot}
\dot{\nu}\simeq \dot\nu_0(1+2 k(\sin(\phi+2\pi\frac{t}{T})-\sin
\phi))+\ddot\nu_{\rm L}(t-t_0),
\end{equation}
where $\dot\nu_0=\dot\nu(t_0)$, $\ddot\nu_{\rm
L}=-2\alpha\dot\nu_0/t_0$ describes the long-term monotonic
variation of $\dot\nu(t)$. Therefore Equation~(\ref{vdot}) can be
tested with the long-term monitoring observations of individual
pulsars. If the long-term observed average of $\ddot\nu$ is
approximately given by the expression for $\ddot\nu_{\rm L}$ above
(i.e. $\langle\ddot\nu\rangle\simeq\ddot\nu_{\rm L}$), then we can
use the previously reported $\ddot\nu$ obtained from the timing
solution fits of the whole data span as an estimate of
$\ddot\nu_{\rm L}$.

Similarly we also find (Paper I)
\begin{equation}\label{vddot}
\ddot{\nu}\simeq -2\dot{\nu}(\alpha/t_{\rm age}+f C(t)),
\end{equation}
where $t_{\rm age}$ is the real age of the pulsar, $f\equiv 2\pi
k/T$ for the dominating oscillating component, and
$C(t)=\cos(\phi+2\pi \frac{t}{T})$. For relatively young pulsars
with $t_{\rm age}<3\times 10^5$~yr, the first term in
Equation~(\ref{vddot}) dominates and we should have $\ddot{\nu}>0$
if $\alpha>0$. Considering that the characteristic ages ($\tau_{\rm
c}$) of young pulsars are normally several times larger than $t_{\rm
age}$, Equation~(\ref{vddot}) thus explains naturally the observed
$\ddot\nu >0$ for most young pulsars with $\tau_{\rm c}<10^6$ yr.

Without other information about $t$, we replace it with the magnetic
age of a pulsar $t=t_0(B_0/B)^{(1/\alpha)}$ in Equation
(\ref{vddot}); we then have,
\begin{equation}\label{ddot_p2}
\ddot{\nu}\simeq \eta(-\dot{\nu})^{1+\beta}/\nu^{3\beta}+
2\dot{\nu}fC(t),
\end{equation}
where $\beta=1/2\alpha$, $\eta=(3.3\times10^{19}/B_0)^{2\beta}/\beta
t_0$ and $B=3.3\times10^{19}\sqrt{P\dot{P}}$~G is assumed. Thus the
model predicts a correlation between $\ddot{\nu}$ and
$(-\dot{\nu})^{1+\beta}/\nu^{3\beta}$ for young pulsars with
$\tau_{\rm c}<10^6$ yr. Similarly for much older pulsars, the second
term in Equation~(\ref{vddot}) dominates, in agreement with the
observational fact that the numbers of negative and positive
$\ddot\nu$ are almost equal for the old pulsars.

From Equation~(\ref{braking law2}), we can also obtain,
\begin{equation}
n_{\rm b}=3+\frac{\tau_{\rm c}}{t}(2- 4 ft C(t)).\label{n_b}
\end{equation}
As we have shown in Paper I, the oscillatory term can be ignored in
determining $\ddot{\nu}$, so young pulsars with $\tau_{\rm c}\le
10^5$ should have $n_{\rm b}>0$, consistent with observations.

\section{Dipole magnetic field evolution of individual pulsars}

\begin{figure}[h]
\center{
\includegraphics[angle=0,scale=0.3]{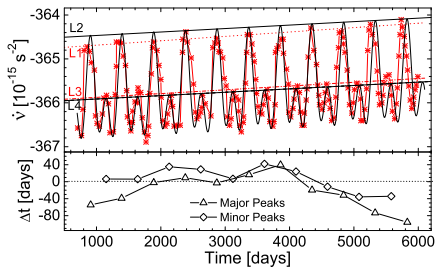}}
\caption{{\it Upper Panel}: $\dot\nu (t) $ for PSR B1828$-$11 during
the past 20 years. The reported data are represented by red stars;
and the solid black line is calculated from Equation~(\ref{vdot}).
{\it Lower Panel}: Time differences between the peak positions of
reported data and analytical calculation. The figure is taken from
Paper V.} \label{Fig:2}
\end{figure}
In Equation (\ref{vdot}), we found that $\dot\nu$ evolution contains
a long-term change modulated by short-term oscillations. It is very
interesting to check whether the long-term changes can be unveiled
from the observational data of some individual pulsars. The sample of Lyne et al. (2010) provides the precise histories of
$\dot\nu$ for seventeen pulsars and thus may be applied to test
Equation~(\ref{vdot}). In the sample, the $\dot\nu$ evolutions for
most of the pulsars exhibit complex patterns. A subset of pulsars
with small $\tau_{\rm c}$ and $\tau_{\rm B}$ are thus selected to
reveal clearly their long-term magnetic field changes.

Figure \ref{Fig:2} (taken from Paper V) shows the comparison between
the reported and analytically calculated $\dot\nu(t)$ for B1828-11.
The one major difference is caused by the decrease of the
oscillation periods of the reported data after $\sim4000$~days.
Nevertheless, our model describes the general trend of the reported
data quite well. From the results, we found $\ddot\nu_{\rm L}>0$ for the pulsar,
which means that $\alpha>0$, i.e., magnetic field decay is directly
observed for them, as predicted by our phenomenological model. The
decay time scale is $|\tau_{B}|=3.3\times 10^4~{\rm yr}$. The
alternative possibility that it is caused by the magnetic
inclination change is ruled out with the data of the position angle
and pulse width changes (Paper V). Theoretically, there are three
avenues for magnetic field decay in isolated NSs, ohmic decay,
ambipolar diffusion, and Hall drift (Goldreich \& Reisenegger 1992).
We found that the Hall drift at outer crust of the NS is responsible
for the field decay, which gives a time scale
$|\tau_{\rm Hall}|=1.1\times 10^4~{\rm yr}$ that agrees with $\tau_{\rm
B}$ of the pulsar (see Paper V for the details). The time scales for
the other avenues are too long and thus not important. The
consistency between the two time scales also implies that the
majority of dipolar magnetic field lines are restricted to the outer
crusts (above the neutron drip point), rather than penetrating the
cores of the NSs.

The diffusive motion of the magnetic fields perturbs the background
dipole magnetic fields at the base of the NS crust. Such
perturbations propagate as circularly polarized ``Hall waves" along
the dipole field lines upward into the lower density regions in the
crusts. The Hall waves can strain the crust, and the elastic
response of the crust to the Hall wave can induce an angular
displacement (Cumming et al. 2004)
\begin{equation}\label{Strain}
\theta_{\rm s}=3\times 10^{-7} B_{12}^2n^{13/9}\frac{\delta B_{\rm
b}}{B},
\end{equation}
in which $B_{12}$ is the strength of the dipole magnetic fields with
unit of $10^{12}$ G, $n$ is the wave number over the crust, and
$\delta B_{\rm b}$ is the amplitude of the mode at the base of the
crust. We found that the short-term oscillations in $\dot\nu$ and
pulse width can be explained dramatically well with moderate values
of the parameters, $n=1200$ and $\delta B/B=0.2$ (Paper V).

Therefore, we concluded that the Hall drift and Hall waves in NS
crusts are responsible for the observed long-term evolution of the
spin-down rates and their quasi-periodic modulations, respectively.

\section{Statistical properties of pulsar timing noise}

\subsection{Reproducing the observed distribution of $\ddot\nu$ and $n_{\rm b}$}

Hobbs et al. (2010; hereafter H2010) carried out a very extensive
study on observed $\ddot\nu$ for 366 pulsars. Some of their main
results are: (1) All young pulsars have $\ddot{\nu} > 0$; (2)
Approximately half of the older pulsars have $\ddot{\nu} > 0$ and
the other half have $\ddot{\nu} < 0$; and (3) The value of
$\ddot{\nu}$ measured depends upon the data span and the exact
choice of data processed. In Figure \ref{Fig:1} (which is taken from
Paper I), we plotted the comparison between the observed $\ddot \nu$
and that predicted by Equation ({\ref{ddotnu}}). This model predicts
$\ddot \nu>0$, against the fact that many pulsars show $\ddot
\nu<0$. This is a major failure of this model. It is also clear that
this model under-predicts the amplitudes of $\ddot \nu$ by several
orders of magnitudes for most pulsars.

In Figure~\ref{Fig:2}(a, b) (Paper I), we show the
comparison between the predicted correlation between $\ddot{\nu}$
and $(-\dot{\nu})^{1+\beta}/\nu^{3\beta}$ for young pulsars with
$\tau_{\rm c}<10^6$ yr by Equation~(\ref{ddot_p2}) and data with
$\alpha=0.5$ and $\alpha=1.0$, respectively. In both cases the model
can describe the data reasonably well. We thus conclude that a
simple power-law decay model is favored by data.

\begin{figure}[h]
\center{
\includegraphics[angle=0,scale=0.35]{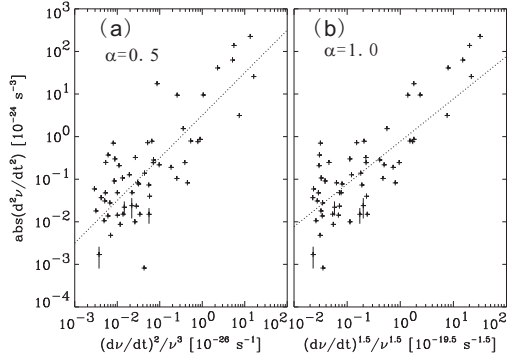}}
\caption{Correlations between $\ddot{\nu}$ and
$(-\dot{\nu})^{1+\beta}/\nu^{3\beta}$ with $\alpha=0.5$ (panel (a))
and $\alpha=1.0$ (panel (b)) for young pulsars with $\tau_{\rm
c}\leq 2\times 10^6$~yr and $\ddot \nu>0$. The dotted lines are the
best-fit of $\ddot{\nu}=\eta(-\dot{\nu})^{1+\beta}/\nu^{3\beta}$.
This figure is taken from Paper I.} \label{Fig:2}
\end{figure}

\begin{figure}[h]
\center{
\includegraphics[angle=0,scale=0.45]{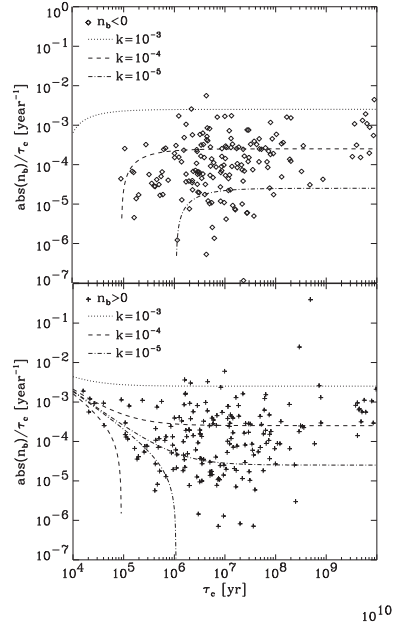}}
\caption{Correlation between $n_{\rm b}/\tau_{\rm c}$ and $\tau_{\rm
c}$. The different curves show the analytically calculated model
predictions using Equation~(\ref{n_b}) for different values of $k$.
The figure is taken from Paper II.} \label{Fig:4}
\end{figure}

In Figure~\ref{Fig:4} (taken from Paper II), we show the observed
correlation between $n_{\rm b}/\tau_{\rm c}$ and $\tau_{\rm c}$,
overplotted with the analytical results of Equation~(\ref{n_b}) with
$C(t)=\pm 1$, $T=10$~yr and $k=10^{-3},\ 10^{-4}$ and $10^{-5}$,
respectively. Once again, the analytical results agree with the data
quite well.

\begin{figure}
\center{
\includegraphics[angle=0,scale=0.2]{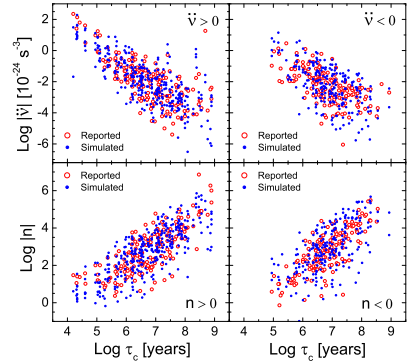}}
\caption{$\ddot\nu$-$\tau_{\rm c}$ and $n$-$\tau_c$ diagrams. The
simulated data and reported data are represented with solid circles
and open circles, respectively. The figure is taken from Paper IV.}
\label{Fig:5}
\end{figure}

We also performed Monte Carlo simulations for the distributions of
reported data (H2010) in $\ddot\nu$-$\tau_{\rm c}$ and $n$-$\tau_c$
diagrams, as shown in Figure \ref{Fig:5} (taken from
Paper IV), respectively. The two dimensional K-S tests show that the
distributions of two samples in each panel are remarkably
consistent.

\begin{figure}[h]
\center{
\includegraphics[angle=0,scale=0.35]{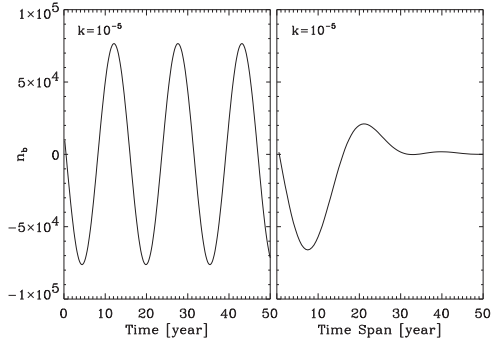}}
\caption{Left: ``instantaneous" braking index $n_{\rm b}$ as a
function of time. Right: ``averaged" braking index $n_{\rm b}$ as a
function of the time span of the fitting. $T=15$~yr is used in both
cases. The figure is taken from Paper II.} \label{Fig:6}
\end{figure}

\subsection{The instantaneous and averaged values of $n_{\rm b}$}

Equation (\ref{n_b}) gives $n_{\rm b}$ as a function of $t$, i.e.,
the calculated $n_{\rm b}$ is in fact a function of time for a given
pulsar, as shown in the left panel of Figure~\ref{Fig:6}, in which
the horizontal axis ``Time" is the calendar time. We call $n_{\rm
b}$ calculated this way the ``instantaneous" braking index. However,
in analyzing the observed timing data of a pulsar, one usually fits
the data on TOAs over a certain time span to Equation (\ref{phase}),
where $\Phi (t)$ is the phase of TOA of the observed pulses, and
$\Phi_0$, $\nu_0$, $\dot \nu_0$ and $\ddot\nu_0$ are the values of
these parameters at $t_0$, to be determined from the fitting.
$n_{\rm b}$ calculated from $\nu_0$, $\dot \nu_0$ and $\ddot\nu_0$
is thus not exactly the same as the ``instantaneous" braking index.
We call $n_{\rm b}$ calculated this way over a certain time span the
``averaged" braking index.

In the right panel of Figure~\ref{Fig:6}, we show the simulated
result for the ``averaged" braking index as a function of time span.
It can be seen that the ``averaged" $n_{\rm b}$ is close to the
``instantaneous" one when the time span is shorter than $T$, which
is the oscillation period of the magnetic fields. The close match
between our model predicted ``instantaneous" $n_{\rm b}$ and the
``averaged" $n_{\rm b}$, as shown in Figure~\ref{Fig:4}, suggests
that the time spans used in the H2010 sample are usually smaller
than $T$.

For some pulsars the observation history may be longer than $T$ and
one can thus test the prediction of Figure~\ref{Fig:6} with the
existing data. In doing so, we can also obtain both $f$ and $T$ for
a pulsar, thus allowing a direct test of our model for a single
pulsar. We can in principle then include the model of magnetic field
evolution for each pulsar in modeling its long term timing data, in
order to remove the red noise in its timing residuals, which may
potentially be the limiting factor to the sensitivity in detecting
gravitational waves with pulsars.

\section{A phenomenological model for glitches}

In this section, we describe the phenomenological spin-down model
for the glitch and slow glitch recoveries (see Xie \& Zhang 2013;
hereafter Paper III). We found that Equation~(\ref{braking law2})
can be modified slightly to describe a glitch event,
\begin{equation}\label{rredipole}
\dot\nu\nu^{-3} =-H_0 G(t),
\end{equation}
where $H_0=\frac{8\pi^2(BR^3\sin\chi)^2}{3c^3I}=1/2\tau_{\rm
c}\nu_0^2$, $\tau_{\rm c}=-\nu/2\dot\nu$ is the characteristic age
of a pulsar, and $G(t)$ represents very small changes in the
effective strength of dipole magnetic field $B\sin\chi$ during a
glitch recovery. In the following we assume $G(t)=1+\kappa
e^{-\Delta t/\tau}$.

Integrating and solving Equation~(\ref{rredipole}), we have
\begin{equation}\label{nu}
\nu(t)\approx \nu_0+\Delta\nu_{\rm d}e^{-\Delta t/\tau}.
\end{equation}
The derivative of $\nu$ is
\begin{equation}\label{dnu}
\dot\nu(t)\approx \dot\nu_0-\Delta\dot\nu_{\rm d}e^{-\Delta t/\tau}.
\end{equation}
We know $\Delta t\sim \tau \sim 100~{\rm days}$ and $\kappa\ll 1$.

\begin{figure}
\centering
\includegraphics[scale=0.35]{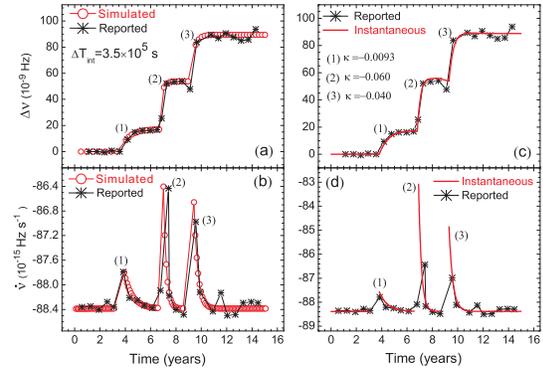}
\caption{Slow glitches of Pulsar B1822--09. Observational results
are taken from Shabanova (2005). Upper panels: variations of
$\Delta\nu$ relative to the pre-glitch solution. Bottom panels:
variations of $\dot{\nu}$. Left panels: comparison between the
reported and simulated (both are also time-averaged) $\Delta\nu$ and
$\dot\nu$. Right panels: comparison between the reported and
restored (i.e. model-predicted) instantaneous $\Delta\nu$ and
$\dot\nu$. The figure is taken for Paper III.} \label{Fig:7}
\end{figure}

In Figure {\ref{Fig:7}} (taken from Paper III), we show the
simulations for the reported three slow glitches of B1822-09 over
the 1995-2004 interval. We confirmed that the slow glitch behavior
can be explained by our phenomenological model with $\kappa<0$. It
is also clear that the instantaneous values of $\Delta\dot\nu$,
which are obtained directly with the model with the parameters are
given by the simulation, are much larger than the reported results
in literature.

\begin{figure}
\centering
\includegraphics[angle=0,scale=0.32]{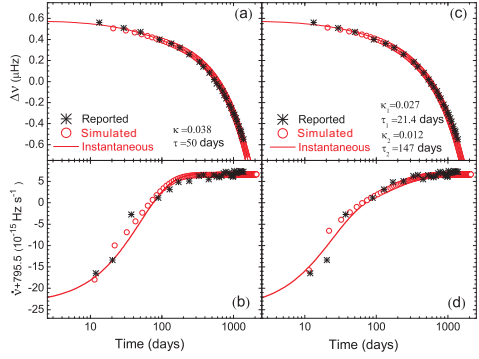}
\caption{The giant glitch of Pulsar B2334+61. Observational results
are taken from Yuan et al. 2010. Upper panels: variations of
$\Delta\nu$. Bottom panels: variations of $\dot{\nu}$. The left and
right panels represent for models with one and two decay components,
respectively. The figure is taken for Paper III.} \label{Fig:8}
\end{figure}

Yuan et al. (2010) reported a very large glitch occurred between
2005 August 26 and September 8 (MJDs 53608 and 53621), the largest
known glitch ever observed, with a fractional frequency increase of
$\Delta\nu/\nu\sim20.5\times10^{-6}$. In the left panels of
Figure~\ref{Fig:8}, we show the fits with one exponential term
$G(t)=(1+\kappa\exp{(-\Delta t/\tau)})$ for a comparison with the
``realistic'' simulation of two terms below. We show the modeled
glitch recovery with $G(t)=(1+\kappa_1\exp{(-\Delta
t/\tau_1)}+\kappa_2\exp{(-\Delta t/\tau_2)})$ in the right panels of
Figure~\ref{Fig:8}. Clearly the simulated profiles of the two term
fit matches the reported ones better than that of the one term fit.
One can see that $|\Delta \dot\nu_{\rm I}|$ are also slightly larger
than the reported $|\Delta \dot\nu_{\rm O}|$ for both the one-term
fit and two-term fit.

\begin{figure}
\centering
\includegraphics[scale=0.32]{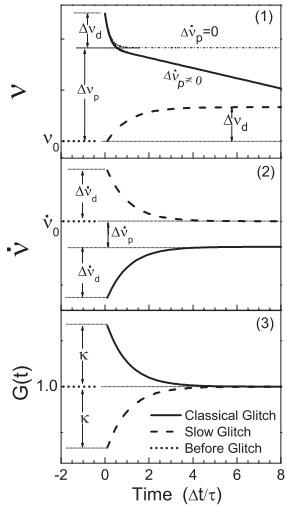}
\caption{Schematic depictions of $\nu$, $\dot\nu$ and $G(t)$ for the
slow and classical glitch recoveries. The pre-glitch tracks are
represented by dotted line. The classical glitch recoveries are
represented by solid lines. The slow glitches are represented by
dashed lines.} \label{Fig:9}
\end{figure}

We thus concluded that the classical and slow glitch recoveries can
be well modeled by a simple function, $G(t)=1+ \kappa\exp{(-\Delta
t/\tau)}$, with positive or negative $\kappa$, respectively. Based
on the results, we generalize the variations of $\nu$ and $\dot\nu$
for slow and classical glitch recoveries, as shown in
Figure~\ref{Fig:9}. The pre-glitch tracks are represented by dotted
line. After the jump, the classical glitch recoveries (represented
by solid line) generally have $\nu$ variation that tends to restore
its initial values, and usually the restoration is composed by a
exponential decay and a permanent linear decrease with slope
$\Delta\dot\nu_{\rm p}$; however, for slow glitches (represented by
dashed line), $\nu$ monotonically increases, as shown in panel (1).
In panel (2), $\dot\nu$ of classical glitch recoveries that tends to
restore its initial values, but cannot completely recover for
$\Delta\dot\nu_{\rm p}\neq 0$; $\dot\nu$ of slow glitch recoveries
almost completely recover to its initial value, corresponding to the
increase of $\nu$.

The function, $G(t)=1+ \kappa\exp{(-\Delta t/\tau)}$, with positive
or negative $\kappa$, are shown in panel (3), respectively. However,
it is should be noticed that the model only have two parameters,
$\kappa$ and $\tau$, from which we can obtain $\Delta\nu_{\rm d}$
and $\Delta\dot\nu_{\rm d}$, but not $\Delta\nu_{\rm p}$ and
$\Delta\dot\nu_{\rm p}$, which are not modelled. The expression of
$\Delta\nu_{\rm p}$ and $\Delta\dot\nu_{\rm p}$ that relate to the
initial jumps of $\nu_0$ and $\dot\nu_0$, are not given by the
model, since the glitch relaxation processes are only considered
here. It has been suggested that these non-recoverable jumps are the
consequence of permanent dipole magnetic field increase during the
glitch event (Lin and Zhang 2004). Nevertheless, we conclude that
the major difference between slow glitch and classical glitch
recoveries are that they show opposite trends with opposite signs of
$\kappa$, in our phenomenological model.

Also as shown above, all the reported results of all pulsar glitches are systematically biased, and thus cannot be compared directly with theoretical models.
In Paper III, we carried extensive simulations in examining all possible sources of biases due to imperfections both in observations, analysis methods, and the ways in reporting results in literature. We suggested some fitting procedures that can significantly reduce the
biases for fitting the observed glitch recoveries and comparison with theoretical models.

\section{Summary}

We tested models of magnetic field evolution of NSs with the observed spin evolutions of individual pulsars and the
statistical properties of their timing noise. In
all models, the magnetic dipole radiation is assumed to dominate the instantaneous
spin-down of pulsars; therefore, different models of their magnetic
field evolution lead to different properties of their spin-down. We
constructed a phenomenological model of the evolution of the
magnetic fields of NSs, which is made of a long-term decay modulated
by short-term oscillations. By comparising our model predictions with the precisely observed spin-down evolutions of some individual pulsars, we found that
the Hall drift and Hall waves in the NS crusts are responsible for
the long-term change and short-term quasi-periodical oscillations,
respectively. We showed that the observed braking
indices of the pulsars in the sample of H2010, which span over a
range of more than 100 millions, can be completely reproduced with
the model. We find that the ``instantaneous" braking index of a
pulsar may be different from the ``averaged'' braking index obtained
from data. We also presented a phenomenological model for the
recovery processes of classical and slow glitches, which is used to
successfully model the observed slow and classical glitch events
from pulsars B1822-09 and B2334+61, respectively. Significant biases are found for fitting glitch recovery, in the widely used fitting procedures and reported results in literature.

\acknowledgements  SNZ acknowledges partial funding support by 973 Program of China under grant Nos. 2009CB824800 and 2014CB845802, by the National Natural Science Foundation of China under grant Nos. 11133002 and 11373036, and by the Qianren start-up grant 292012312D1117210.

\end{document}